\documentclass[aps,twocolumn,amsmath,amssymb,showpacs,nofootinbib]{revtex4-1}
\usepackage{bm}

\usepackage{latexsym}
\usepackage{amsthm}
\usepackage{amsmath}
\usepackage{braket}
\usepackage{mathtools}
\usepackage[T1]{fontenc}
\allowdisplaybreaks

\usepackage[top = 1. in,bottom = .9 in, left = 1 in, right=1 in]{geometry}

 \newcommand{\arXiv}[1]{\href{http://www.arXiv.org/abs/#1}{arXiv:#1}}
\usepackage[colorlinks=true, linkcolor=blue, bookmarks=true]{hyperref}

\makeatletter
\renewcommand\section{\@startsection {section}{1}{\z@}%
                  {-3.5ex \@plus -1ex \@minus -.2ex}
                  {2.3ex \@plus.2ex}%
                  {\normalfont\bfseries}}
\renewcommand\subsection{\@startsection{subsection}{2}{\z@}%
                   {-3.25ex\@plus -1ex \@minus -.2ex}%
                   {1.5ex \@plus .2ex}%
                   {\normalfont\bfseries}}
\makeatother

\newcommand{\al}{\alpha}

\newcommand{\beq}{\begin{equation}}
\newcommand{\eeq}{\end{equation}}
\newcommand{\ber}{\begin{array}}
\newcommand{\eer}{\end{array}}

\newcommand{\dsty}{\displaystyle}

\newcommand{\de}{\delta}

\newcommand{\eps}{\varepsilon}

\newcommand{\om}{\omega}

\newcommand{\ena}{\end{eqnarray}}
\newcommand{\beqa}{\begin{eqnarray}}
\newcommand{\eeqa}{\end{eqnarray}}
\newcommand{\bea}{\begin{eqnarray}}
\newcommand{\eea}{\end{eqnarray}}

\theoremstyle{remark}

\begin{document}

\title{Time-periodic quantum states of weakly interacting bosons in a harmonic trap}
\author{Marine De Clerck$^{1}$ and Oleg Evnin$^{2,1}$\vspace{2mm}}

\affiliation{ $^{1}$Theoretische Natuurkunde, Vrije Universiteit Brussel (VUB) and\\
	The International Solvay Institutes, Pleinlaan 2, B-1050 Brussels, Belgium\vspace{1mm}\\
$^{2}$Department of Physics, Faculty of Science, Chulalongkorn University, Phayathai Rd., Bangkok 10330, Thailand}

\begin{abstract}
We consider quantum bosons with contact interactions at the Lowest Landau Level (LLL) of a two-dimensional isotropic harmonic trap. At linear order in the coupling parameter $g$, we construct a large, explicit family of quantum states with energies of the form $E_0+gE_1/4+O(g^2)$, where $E_0$ and $E_1$ are integers. Any superposition of these states evolves periodically with a period of $8\pi/g$ until, at much longer time scales of order $1/g^2$, corrections to the energies of order $g^2$ may become relevant. These quantum states provide a counterpart to the known time-periodic behaviors of the corresponding classical (mean field) theory.
\end{abstract}

\maketitle

\section{Introduction}

Systems of identical interacting bosons in a harmonic trap are a standard topic in the physics of cold atomic gases \cite{BDZ}. While much of the effort in this domain has been given to the emergence of Bose-Einstein condensation, the more basic and general question of the quantum energy spectra and the corresponding wavefunctions in the presence of interactions has likewise attracted a considerable amount of attention \cite{PR, split1, split2,split3,split4,split5,split6, split7, split8,split9,split10}.

In our previous article \cite{split}, we systematically addressed the structure of fine splitting of the energy levels in a two-dimensional isotropic harmonic trap due to weak contact interactions at linear order in the coupling parameter \cite{renormalons}. Our goal has been to explore the operation of the symmetries of the system with respect to the diagonalization problems defining the level splitting, to describe optimizations in the computation of the spectra due to the symmetry structure, to establish bounds on the ranges of the spectra, and to investigate, in the spirit of the quantum chaos theory \cite{QCh}, the energy level statistics within the interaction-induced fine level splitting, in order to identify, or rule out, additional symmetry structures.

In this article, we shall present a more refined treatment of the energy levels within the Lowest Landau Level (LLL), identifying large families of explicit energy eigenstates. Particles in LLL states have the maximal amount of angular momentum for a given energy, and LLL restrictions have often been studied in relation to rapidly rotating Bose-Einstein condensates \cite{ho,LLLvortex,fetter}. With respect to the fine structure splitting, the analysis of the LLL levels completely separates from all other levels, as explained in \cite{split}, and can be treated independently.

Our specific attention to the LLL levels is motivated by the known peculiarities of the corresponding classical problem. Indeed, the splitting of the energy levels at linear order in the coupling parameter is governed by the resonant LLL Hamiltonian \cite{split}. The classical dynamics of this Hamiltonian has been investigated in a few works \cite{continuous, BBCE, GGT}, driven by rather different motivations. Explicit classical solutions are known for the LLL Hamiltonian, and these known solutions have mode amplitude spectra exactly periodic in time \cite{BBCE}. One can then ask what features of the quantum energy eigenstates underlie these simple, explicitly known classical behaviors.

What we find is much more than what we ask for. While in principle, the time-periodic properties of the classical orbits could emerge from asymptotic properties of the quantum energy eigenstates at high energies, without any corresponding exact periodicities in the quantum theory, what we find is that there are ladders of equispaced energy levels at weak coupling, with distances between such ladder state energies given by a fixed number times the coupling constant. The wavefunctions of the ladder states can be constructed explicitly, and while these states form a small subset of all energy eigenstates, their number grows without bound as the number of particles increases. These states underlie the classical periodic behaviors, to which they can be explicitly connected by forming coherent-like combinations, and in fact provide a much richer array of periodic behaviors than what is known to exist in the classical theory.

While the derivations in our paper will be given for nonrelativistic bosons common in the physics of cold atomic gases, closely related structures arise in other branches of physics. Relativistic analogs of the systems we study exist in Anti-de Sitter spacetimes, displaying classical features closely reminiscent of the LLL Hamiltonian \cite{CF,BMP,BEL,BEF}. These features can be distilled into a specific class of partially solvable resonant Hamiltonians \cite{AO}, and are rooted in the existence of breathing modes \cite{breathing}, similar to the Pitaevskii-Rosch breathing mode \cite{PR} or to the exactly periodic center-of-mass motion \cite{bbbb} in our present system. Quantum energy level corrections due to interactions in Anti-de Sitter spacetimes, which are directly parallel to the nonrelativistic topics of our current paper, have attracted attention within high-energy theory \cite{BSS,madagascar}. Generalizations of our current findings to these relativistic cases (which we shall not immediately pursue here), will provide new results in that domain of study.

We shall now proceed with presenting the concrete results of our investigations. We will first review, in the next section, the basics of weakly nonlinear energy level splitting for bosons with contact interactions in a two-dimensional harmonic trap. In section \ref{sec: Ladder states},  we shall present an explicit construction of families of eigenstates with equidistant energies, so that arbitrary superpositions of such states evolve
periodically in time. In section \ref{sec:coherent}, we shall present coherent-like combinations of these states that have an explicit connection to time-periodic solutions of the classical weakly nonlinear theory. We shall conclude with a discussion, highlighting possible generalizations, and the way our quantum considerations elucidate, in a nontrivial way, the behavior of the corresponding classical theory.


\section{Weakly interacting bosons\\ at the lowest Landau level}

We shall give a very brief review of the energy level splitting for bosons in a harmonic trap due to weak contact interactions. Further details can be found in \cite{split}.

\subsection{General energy levels}

Our main object of study is the following second-quantized Hamiltonian:
\begin{align}
 &\mathcal{H} =\mathcal{H}_0 + g \,\mathcal{H}_{int},  \label{Horg}\\
&\mathcal{H}_0= \frac{1}{2}\int (  \nabla \Psi^\dagger \cdot \nabla \Psi + (x^2 + y^2)\Psi^\dagger \Psi)\,dx\,dy,\nonumber\\ 
&\mathcal{H}_{int}=\pi \int \,  \Psi^{\dagger2} \Psi^{2} \, dx\,dy,\nonumber
\end{align}
where the interaction strength $g$ is taken to be small, $0<g\ll 1$. (The restriction to positive $g$ is completely inessential, but simply adopted to simplify the wording in our considerations.)
The field operators $\Psi(x,y)$ and $\Psi^\dagger(x,y)$ satisfy the standard commutation relations
\beq
[\Psi^\dagger(x,y),\Psi(x',y')]=-\de(x-x')\,\de(y-y').
\eeq
There are known subtleties in defining contact interactions discussed, for instance, in \cite{split1}, but they play no role in our treatment\footnote{These subtleties, however, become crucial if one studies high orders of the perturbative expansion, rather than the lowest order. For some related systems with contact interactions, fascinating structures emerge through such studies in relation to Borel summability of the perturbative series, see  \cite{renormalons}.} as we restrict our attention to the linear order in $g$ \cite{split}.

In a noninteracting theory with $g=0$, the bosons independently occupy the energy levels in the harmonic potential, and the total energy is the sum of the individual (integer) energies. The field operator $\Psi$ can be decomposed in terms of the harmonic oscillator eigenfunctions $\psi_{nm}$ carrying  $n+1$ units of energy and $m$ units of angular momentum (one must have $m \in \{-n, -n+2, \dots, n-2,n\}$):
\beq
\Psi(x,y)=\sum_{n,m} a_{nm} \psi_{nm}(x,y).
\label{psidecomp}
\eeq
The creation-annihilation operators $a^\dagger_{nm}$ and $a_{nm}$ satisfy
\beq
[a^\dagger_{nm},a_{n'm'}]=-\de_{nn'}\de_{mm'}.
\eeq
The Hamiltonian $\mathcal{H}_0$ is diagonal in the Fock basis generated by these operators. To obtain a state with occupation numbers $\{\eta_{nm}\}$ from the vacuum $|0\rangle$, one writes
\beq
|\{\eta_{nm}\}\rangle=\prod_{nm}\frac{( a^\dagger_{nm} )^{\eta_{nm}}}{\sqrt{\eta_{nm}!}}|0\rangle.
\label{Fockdef}
\eeq
The corresponding energies (minus the vacuum energy) are then simply
\beq
E_{\{\eta\}}=\sum_{nm} n \,\eta_{nm}.
\label{E0def}
\eeq
The degeneracies of these energy levels are very high, given by the number of ways to partition a given integer $E$ by populating different modes with particles. The degeneracies grow without bound at higher values of $E$.

Turning on weak interactions, one can write
\begin{align}
\mathcal{H}_{int}=&\hspace{2mm}{\textstyle\frac12} \,\sum_{\mathclap{\substack{n_1,n_2,n_3,n_4 \geq 0\\m_1+m_2 = m_3+ m_4}}} \hspace{2mm}C_{n_1n_2n_3n_4}^{m_1m_2m_3m_4}  {a}^{\dagger}_{n_1m_1}{a}^{\dagger}_{n_2m_2} {a}_{n_3m_3}{a}_{n_4m_4}.
\label{eq: interaction Hamiltonian}
\end{align}
The condition $m_1+m_2 = m_3+ m_4$ is dictated by the angular momentum conservation. We have furthermore introduced the \emph{interaction coefficients} $C_{n_1n_2n_3n_4}^{m_1m_2m_3m_4}$, which are a set of numbers computed from integrals of products of $\psi_{nm}$ \cite{split}. To compute the energy shifts at linear order in $g$, one simply has to evaluate all matrix elements
\beq
\langle \{\eta\}|\mathcal{H}_{int}| \{\eta'\}\rangle
\label{blockdef}
\eeq 
between Fock states $| \{\eta\}\rangle$ and $| \{\eta'\}\rangle$ with the same energy $E$, same number of particles $N=\sum_{nm} \eta_{nm}$ and the same angular momentum $M=\sum_{nm} m\,\eta_{nm}$.
For any given $N$, $E$ and $M$, this is a finite-sized matrix whose eigenvalues $\eps_I$ give the energy shifts in the fine structure of the original energy level $E$ at order $g$:
\beq
\tilde E_I=E+g\,\eps_I.
\label{enshifts}
\eeq

One may notice that only those terms in (\ref{eq: interaction Hamiltonian}) that satisfy $n_1+n_2=n_3+n_4$ may contribute to the matrix elements (\ref{blockdef}), since $| \{\eta\}\rangle$ and $| \{\eta'\}\rangle$ carry the same energy. One can then replace $\mathcal{H}_{int}$ in (\ref{blockdef}) by
\beq
\mathcal{H}_{res}=
{\textstyle\frac12} \, \sum_{\mathclap{\substack{n_1+n_2 = n_3+n_4 \\m_1+m_2 = m_3+ m_4}}}  C_{n_1n_2n_3n_4}^{m_1m_2m_3m_4}  {a}^{\dagger}_{n_1m_1}{a}^{\dagger}_{n_2m_2} {a}_{n_3m_3}{a}_{n_4m_4}.
\label{resH}
\eeq
The classical version of this {\it resonant Hamiltonian} is frequently encountered in studies of weakly nonlinear long-term dynamics of resonant PDEs, see, for instance, \cite{continuous} for applications to the nonlinear Schr\"odinger equation in a harmonic potential, the classical version of (\ref{Horg}).

Besides conserving $N$, $M$ and $E$, the resonant Hamiltonian (\ref{resH}) possesses extra conservation laws whose explicit form can be found in \cite{split}, originating from the breathing modes \cite{breathing} of (\ref{Horg}). These extra conservation laws impose relations between energy shifts in unperturbed energy levels with different values of $N$, $M$ and $E$. A detailed description of the patterns in the fine structure spectra induced by these symmetries can be found in \cite{split}, together with the ways the symmetries can be used for optimizing computations of the spectra. We shall give below an explicit description of these algebraic structures for the LLL levels, the case of interest for us in this article, where the mathematical details simplify and can be stated compactly.

\subsection{The LLL truncation}

The classical dynamics corresponding to (\ref{resH}) can be consistently truncated to the set of modes with the maximal amount of rotation for a given energy, namely, the modes in (\ref{psidecomp}) satisfying $n=m$. Such a truncation has appeared in the literature under the name of the Lowest Landau Level (LLL) equation \cite{continuous, BBCE, GGT}.
One does not in general expect that consistent classical truncations have direct implications in the quantum theory, since quantum variables cannot be simply set to zero. It turns out, however, that there is a precise quantum counterpart of the classical LLL truncation. Namely, one can convince oneself \cite{split} that the only way to have $E=M$ in an unperturbed energy level (\ref{Fockdef}) is if the only nonzero occupation numbers $\eta_{nm}$ are for modes with $n=m$. Thus, states of this type have vanishing matrix elements (\ref{blockdef}) with any other states, while their nonvanishing matrix elements among themselves are completely governed by the part of the resonant Hamiltonian (\ref{resH}) that only depends on the creation-annihilation operators with $n=m$. This part, the LLL Hamiltonian, can be expressed simply as
\begin{align} \label{resonant_LLL}
\mathcal{H}_{LLL} = {\textstyle\frac{1}{2}} \hspace{-5mm}\sum^{\infty}_{\substack{n_1,n_2,n_3,n_4=0\\n_1 +n_2= n_3+ n_4}}\hspace{-5mm}C_{n_1n_2n_3n_4} a^\dagger_{n_1} a^\dagger_{n_2} a_{n_3} a_{n_4}
\end{align}
with
\begin{align}
C_{n_1n_2n_3n_4} = \frac{((n_1+n_2+n_3+n_4)/2)!}{2^{n_1+n_2}\sqrt{n_1!n_2!n_3!n_4!}}.
\label{CLLL}
\end{align}
We have renamed $a_{n_im_i}$ with $n_i=m_i$ to $a_{n_i}$.

The LLL Hamiltonian (\ref{resonant_LLL}-\ref{CLLL}) commutes with the following operators, which reflect a subset of the conserved quantities of (\ref{resH}) that we have already briefly mentioned:
\begin{align}
&{N} = \sum_{k=0}^\infty {a}^\dagger_k {a}_k, \qquad {M} = \sum_{k=1}^\infty k {a}^\dagger_k {a}_k,\nonumber\\
 &{Z} = \sum_{k=0}^\infty \sqrt{k+1} \,a^\dagger_{k+1} a_k.\label{ZLLL}
\end{align}
Physically, $Z$ is a raising operator for the center-of-mass motion (which is an independent harmonic oscillator decoupled from the other degrees of freedom \cite{bbbb}).
The algebra of these conserved quantities is given by
\begin{align}
& [{M}, {Z}] = {Z}, \quad [{M}, {Z}^\dagger]=-{Z}^\dagger,\nonumber\\
& [Z,Z^\dagger]=-N,\label{NEZcomm}
\end{align}
with the remaining commutators vanishing.

The generalities of finding the spectra for systems of the form (\ref{resonant_LLL}), with arbitrary interaction coefficients $C$, have been explained in \cite{quantres}, and are a simplified version of the brief presentation in the previous section (because the set of modes is simpler). We shall now review the construction of the spectra of (\ref{resonant_LLL}-\ref{CLLL}) in more detail because this material forms essential backgrounds for the new derivations we shall present in this article.

The LLL Fock basis is given by
\beq
|\eta_0,\eta_1,\dots\rangle=\prod_{k=0}^\infty\frac{(a^\dagger_k)^{\eta_k}}{\sqrt{\eta_k!}}|0,0,0,\dots\rangle,
\eeq
such that, for any $k$, 
\begin{align}
{a}_k^\dagger {a}_k |\eta_0,\eta_1,\dots\rangle= \eta_k |\eta_0,\eta_1,\dots\rangle,
\end{align}
where $\eta_k$ are nonnegative integers.
The conserved quantities $N$ and $M$ are diagonal in this basis, and the corresponding eigenvalues are
\begin{align}
N = \sum_{k=0}^\infty \eta_k, \quad M = \sum_{k=1}^\infty k \, \eta_k.
\end{align}

Since $N$ and $M$ commute with the Hamiltonian, the diagonalization is performed independently at each value of $N$ and $M$. The number of states with a given value of $N$ and $M$ is the number of integer partitions of $M$ into at most $N$ parts \cite{quantres}, a well-known number-theoretic function denoted as $p_N(M)$. One thus has to diagonalize an explicit $p_N(M)\times p_N(M)$ numerical matrix to find the energy shifts (\ref{enshifts}) at each value of $N$ and $M$.

\begin{figure}
	\centering
	\includegraphics[scale=0.4]{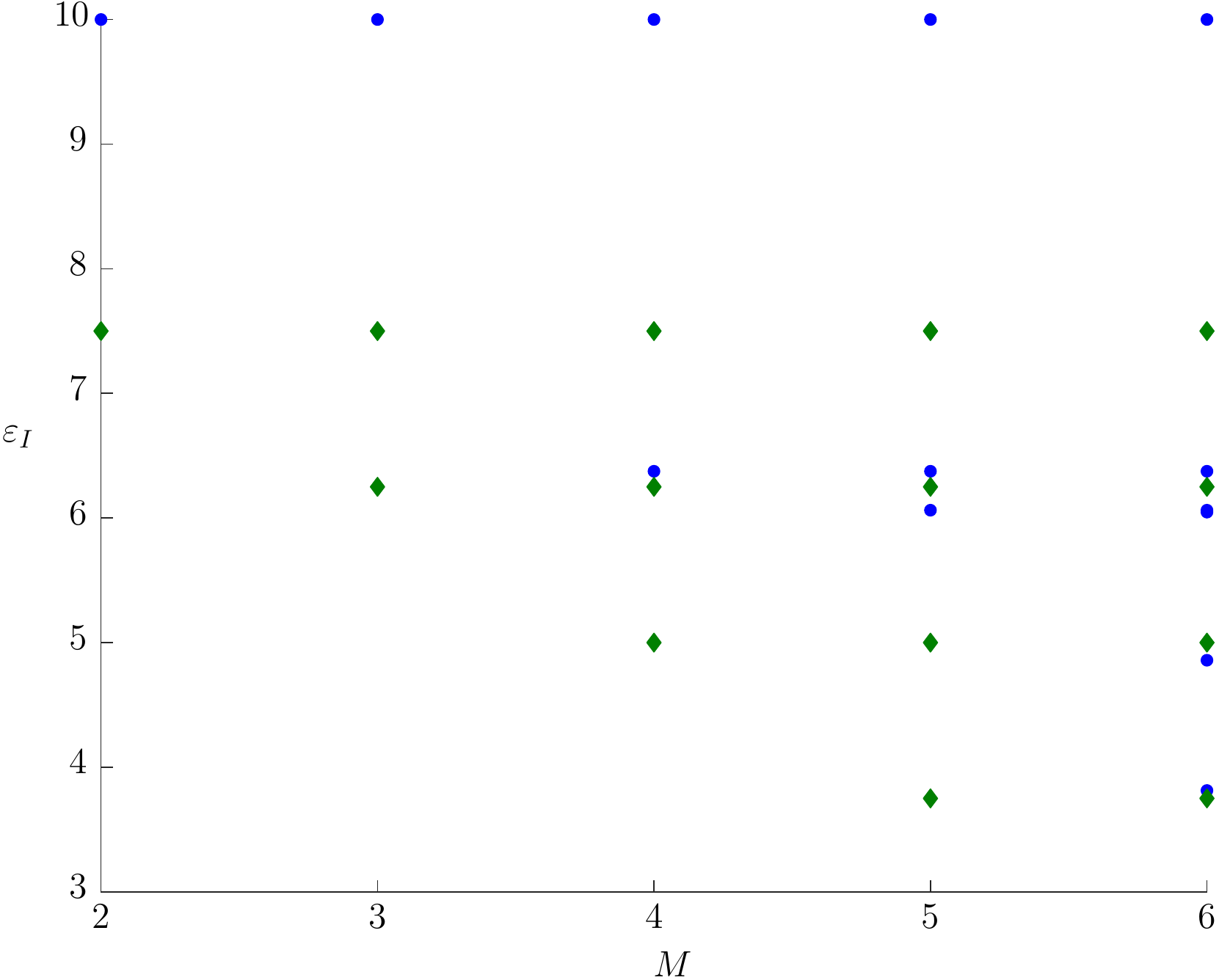}
	\caption{The eigenvalues of $\mathcal{H}_{LLL}$ for a fixed number of particles ($N=5$) as a function of the angular momentum $M$. The eigenvalues at each level ($5,M$) are copied to the next block through the action of the operator $Z$. The eigenvalues corresponding to the ladder states, whose construction is our main goal, are displayed in green (diamonds). For each $M \leq 5$, a single new ladder state eigenvalue appears in the spectrum, which is the lowest eigenvalue at this level.}
    \label{fig: shifts small N}
\end{figure}

The above description is true \cite{quantres} for general values of the interaction coefficients $C$ in (\ref{resonant_LLL}). For the specific values given by (\ref{CLLL}), there is an extra conserved quantity $Z$, which adds a new twist to the story. $Z$ commutes with $\mathcal{H}_{LLL}$, but it acts as a raising operator for $M$ according to (\ref{NEZcomm}). As a result, if $|\Psi\rangle$ is an eigenstate of $\mathcal{H}_{LLL}$ at level $(N,M)$ giving an energy shift $\eps_I$, then $Z|\Psi\rangle$ is an eigenstate of $\mathcal{H}_{LLL}$ at level $(N,M+1)$ with the same eigenvalue $\eps_I$. The action of $Z$ thus copies the energy shifts from levels with lower $M$ to levels with higher $M$, as illustrated in figure~\ref{fig: shifts small N}. In this way, one starts with the vector $|N,0,0,\ldots\rangle$ at level $(N,0)$ satisfying
\beq
\mathcal{H}_{LLL}|N,0,0,\ldots\rangle=\frac{N(N-1)}2 |N,0,0,\ldots\rangle.
\label{N000}
\eeq
The corresponding energy shift $N(N-1)/2$ will evidently be copied by the action of $Z$ to all other blocks with the same value of $N$. This shift, given by $N(N-1)/2$, remains the largest one at all higher levels; it is also true that all eigenvalues of  $\mathcal{H}_{LLL}$ are nonnegative \cite{split}. 

The energy shifts at each level $(N,M)$ are fully exhausted by the energy shifts copied through the action of $Z$ from levels with the same $N$ and lower $M$, and the new shifts that always correspond to energy eigenstates in the kernel of $Z^\dagger$, i.e., states satisfying
\beq
Z^\dagger|\Psi\rangle=0.
\label{Zkern}
\eeq
Indeed, the commutation relation for $Z$ and $Z^\dagger$ \eqref{NEZcomm} implies that the kernel of $Z$ is empty for any nonzero number of particles \cite{split}. As the function $p_N(M)$ is a strictly increasing function for ($M>1$), this means that the dimensionality of the space of states $\left\lbrace \ket{\phi} \right\rbrace$ in the ($N,M+1$)-block that is orthogonal to the image of $Z$ is nonzero. 
However, any such state $\ket{\phi}$ must be annihilated by $Z^\dagger$ since the state $Z^\dagger \ket{\phi}$ belongs to the ($N,M$)-block and is at the same time orthogonal to all states $\left\lbrace\ket{\psi_I}\right\rbrace$ in this block 
\beq
\braket{\psi_I | Z^\dagger \phi} = \braket{ Z \psi_I | \phi} = 0.
\eeq
It is thus in principle enough to study the kernel (\ref{Zkern}) at each $(N,M)$-level to reconstruct all possible shifts at all values of $N$ and $M$, utilizing the action of $Z$ described above.


\section{Ladder states}
\label{sec: Ladder states}

Having spelled out the systematics of diagonalizing the LLL Hamiltonian (\ref{resonant_LLL}-\ref{CLLL}) and its connection to the energy shifts (\ref{enshifts}) at order $g$ for harmonically trapped bosons with contact interactions described by (\ref{Horg}), we proceed with the main result of our paper, namely, the construction of an explicit family of eigenstates of (\ref{resonant_LLL}-\ref{CLLL}), which we call the {\it ladder states}.
In each $Z^\dagger$-kernel (\ref{Zkern}) of an $(N,M)$-level (with $2\le M\le N$), there is precisely one such state with the eigenvalue\footnote{We anticipated such structures based on classical time-periodicities and observed them in numerical simulations, see figure \ref{fig: shifts small N}, before constructing our analytic treatment.}
\beq
\eps^{(N,M)}= \frac{N(N-1)}{2}-\frac{NM}{4}.
\label{epskern}
\eeq
Of course, the action of $Z$ copies such energies from lower levels to higher levels producing, within a given level, a ladder of states with evenly spaced energies of the form
\beq
\eps^{(N,M)}_m= \frac{N(N-1)}{2}-\frac{N(M-m)}{4},
\label{epsladder}
\eeq
where $m$ is any integer number satisfying $M-\min(N,M)\le m\le M-2$. By our discussion above, among such levels, the ones with $m=0$ belong to the kernel (\ref{Zkern}).

Evidently, when the eigenvalues (\ref{epsladder}) are introduced as energy shifts in (\ref{enshifts}), one obtains sets of energy levels of (\ref{Horg}) separated by gaps integer in units of $g/4$, neglecting the higher corrections of order $g^2$.  This would mean that any superposition of the corresponding energy eigenstates will evolve periodically with a period of at most $8\pi/g$, until at very late times of order $1/g^2$ this periodic behavior may be upset by higher-order corrections to the energies.

In the rest of this section, our goal will be to establish the existence of states with energies of the form (\ref{epskern}), and give explicit expressions for their wavefunctions. To this end, we start by examining the following operator
\begin{equation}
B \equiv \mathcal{H}_{LLL} - \frac{N(N-1)}{2} + \frac{1}{4}\left( NM-Z Z^\dagger \right),
\label{eq: definition B}
\end{equation}
which annihilates states with energy shifts given by \eqref{epsladder}.
By (\ref{NEZcomm}), $B$ commutes with both $Z$ and $Z^\dagger$. This particular combination of operators ensures the normal-ordered form of the operator $B$ to be purely quartic in the creation-annihilation operators. 
The precise form of the operator $B$ has been deduced by trial-and-error so as to enable our derivations that follow, with the above algebraic properties of $B$ providing hints, in particular, for the inclusion of the $Z Z^\dagger$ term.
Using (\ref{resonant_LLL}-\ref{CLLL}) and (\ref{ZLLL}), $B$ is itself of the form (\ref{resonant_LLL}) with modified interaction coefficients $C$:
\begin{align}
B &= \mathcal{H}_{LLL}+ \frac14\sum_{n,m=0}^{\infty} (n-2)\, a^\dagger_{n} a^\dagger_{m} a_{n} a_{m} \nonumber
\\
&-\frac{1}{4} \sum_{n,m=0}^{\infty} \sqrt{n+1}\sqrt{m+1}\, a^\dagger_{n+1} a^\dagger_{m} a_{n} a_{m+1}. \nonumber
\end{align}
One can re-express this as
\begin{align}
&B= \frac{1}{2} \sum_{j=0}^{\infty} \sum_{k=0}^{j}A^\dagger_{jk} \Biggl[ \frac{j!}{2^j} \sum_{l=0}^{j} A_{jl}+\frac{k! (j-k)!}4   \nonumber\\
&\times \Big( {\textstyle (j\!-\!4)A_{jk}\!-\!(k\!+\!1)A_{j,k+1}\!-\!(j\!-\!k\!+\!1) A_{j,k-1} }\Big)\Bigg] \nonumber \\
&\hspace{1cm}\equiv \sum_{j=0}^{\infty} \sum_{k,l=0}^{j} B^{(j)}_{kl} A^\dagger_{jk} A_{jl} \label{eq: resonant definition B}, 
\end{align}
where we have defined $A_{jk} \equiv \frac{a_{k}a_{j-k}}{\sqrt{k!(j-k)!}}$. ($A_{j,-1}$ and
$A_{j,j+1}$, which formally occur at the boundaries of the
$k$-summation in (\ref{eq: resonant definition B}), must be understood
as 0.)
Since $A_{jk}$ is invariant under $k\to j-k$, to understand the features of (\ref{eq: resonant definition B}), it suffices to study, for each $j$, the quadratic forms $v^{(j)}_k  B^{(j)}_{kl} v^{(j)}_l$ within the subspaces $v^{(j)}_k=v^{(j)}_{j-k}$.
The following properties of the matrices $(B^{(j)})_{kl}$ within these subspaces deserve to be mentioned:
\begin{enumerate}
\item $(B^{(j)})_{kl}$ vanishes for $j\le 3$.
\item There are two zero eigenvalues corresponding to the eigenvectors
\begin{align}
&v^{(j)}_k=\frac{j!}{k!(j-k)!}\label{zerov}\\ 
\mbox{and}\quad&v^{(j)}_k=\frac{(j-2)!}{(k-1)!(j-k-1)!}.\nonumber
\end{align}
\item All other eigenvalues are positive. (We do not have an analytic proof for this last statement, but we have verified it for a number of different values of $j$ by explicit diagonalization.)
\end{enumerate}
Taken together, these properties imply that $B$ of (\ref{eq: definition B}) is a nonnegative operator:
\beq
\forall|\Psi\rangle:\quad \langle\Psi|B|\Psi\rangle \ge 0.
\label{Bbound}
\eeq
If this bound is saturated,
\beq
B|\Psi\rangle=0.
\label{BPsi}
\eeq
One can explicitly diagonalize $Z Z^\dagger$ using the commutation relations \eqref{NEZcomm} and show that the energies of all states satisfying \eqref{BPsi} are of the form \eqref{epsladder}. 
If, in addition, (\ref{Zkern}) is satisfied, (\ref{epskern}) is ensured, which opens a way to construct the ladder states. We note that the bound (\ref{Bbound}) guarantees that, if a state with the energy (\ref{epskern}) exists within a given $(N,M)$-level, it is the lowest energy state within that $(N,M)$-level. As mentioned above, we do not have a complete proof of this bound, because property 3 in the list has not been proved analytically. This is, however, completely irrelevant for our construction of the ladder states below, since they are explicitly defined as states satisfying (\ref{BPsi}) and (\ref{Zkern}). (The conjectured bound (\ref{Bbound}) ascertains that the energies of the ladder states we find explicitly are the lowest ones within their respective $(N,M)$-levels, which is indeed seen in numerical diagonalization of the LLL Hamiltonian within concrete $(N,M)$-levels, but it is not necessary to construct the ladder states wavefunctions.)

We now proceed with an explicit construction of states with energies given by (\ref{epskern}), for which one needs to solve (\ref{BPsi})  and (\ref{Zkern}) simultaneously.
Property 1 listed above implies that $B$ annihilates the following states 
\begin{align}
&\ket{N-M,M,0,0,0\dots}, \label{nilvec} \\
&\ket{N-M+1,M-2,1,0,0,0\dots}.\nonumber
\end{align}
(Indeed, all terms in $B$ contain either at least one $a_k$ with $k\ge 3$, or $(a_2)^2$, all of which annihilate the above states.)
These states, however, do not satisfy (\ref{Zkern}). One can remedy for that, remembering that $Z$ commutes with $B$, and therefore leaves (\ref{BPsi}) intact, while changing $M$ to $M+1$.
One can therefore take (\ref{nilvec}) at a lower value of $M$, and transport them to the current $(N,M)$-level by repeated action of $Z$. This gives a large set of vectors satisfying (\ref{BPsi}) within the current $(N,M)$-level, and one might hope to form a linear combination of these vectors satisfying both (\ref{BPsi}) and (\ref{Zkern}), yielding the desired ladder state with the energy (\ref{epskern}).

An efficient way to implement in practice the general idea described above is as follows. Consider the set of vectors 
\begin{equation}
\ket{\phi^{NM}_m}=Z^m Z^{\dagger m}\ket{N-M,M,0,\dots},
\label{eq: phiMm}
\end{equation} 
for $m=0 \dots M \leq N$. This is equivalent to considering the states $Z^m \ket{N-M+m,M-m,0,\dots}$ since 
\begin{align}
Z^{\dagger m} &\ket{N-M,M,0,\dots} = \sqrt{\frac{M!(N-M+m)!}{(N-M)!(M-m)!}} \nonumber \\ 
& \, \, \, \, \, \, \times\ket{N-M+m,M-m,0,\dots}.
\label{eq: Zdagger action on O and 1}
\end{align}
As $B$ commutes with $Z$ and $Z^{\dagger}$ and annihilates $\ket{N-M,M,0,\dots}$, all of these vectors solve (\ref{BPsi}). Note also that this set of vectors is linearly independent. This can be understood by noting that $\ket{\phi^{NM}_m}$ contains a term with $a^\dagger_{m+1}$, while all of the $\ket{\phi^{NM}_n}$ with $n < m$ have zero occupation number for this mode and only contain lower modes. Using \eqref{NEZcomm} and \eqref{eq: Zdagger action on O and 1}, we observe that
\begin{align*}
Z^{\dagger} \ket{\phi^{NM}_m} &=  Z^m Z^{\dagger m} Z^\dagger \ket{N-M,M,0,\dots} \\ &+ m N Z^{m-1} Z^{\dagger m-1} Z^\dagger \ket{N-M,M,0,\dots} \\
&\hspace{-1.5cm}= \sqrt{\frac{N-M+1}{M}} \left(\ket{\phi^{N,M-1}_m} + m N \ket{\phi^{N,M-1}_{m-1}}\right).
\end{align*}
We now consider a linear combination of the vectors \eqref{eq: phiMm} at fixed $M$ and $N$:
\begin{align}
\ket{\psi^{N}_M} = \sum^{M}_{m=0} b_m \ket{\phi^{NM}_m},
\end{align}
and determine the coefficients for which this linear combination satisfies (\ref{Zkern}), whose left-hand side is now given by
\begin{align*}
Z^\dagger{\ket{\psi^{N}_M}} &\sim \sum^{M-1}_{m=1} \left( b_m \ket{\phi^{N,M-1}_m}  + mN b_m \ket{\phi^{N,M-1}_{m-1}} \right) \nonumber \\
&\hspace{-1cm}+ b_0 \ket{\phi^{N,M-1}_{0}} +  M N b_M \ket{\phi^{N,M-1}_{M-1}}.
\end{align*}
Because the vectors \eqref{eq: phiMm} are linearly independent, imposing (\ref{Zkern}) gives the following recursion relation
\begin{equation}
b_m = \frac{(-1)^m}{\dsty m! N^m} b_0,
\end{equation}
yielding an explicit (unnormalized) ladder state in the $Z^\dagger$-kernel of level $(N,M)$ in the form
\begin{equation}
\ket{\tilde{\psi}^{N}_M} =  \sum^{M}_{m=0} \frac{(-1)^m}{\dsty m! N^m} Z^m Z^{\dagger m}\ket{N-M,M,0...}.
\label{eq: unnormalized ladder states}
\end{equation}
By construction, this state satisfies (\ref{BPsi}) and (\ref{Zkern}), and hence it is an eigenstate of (\ref{resonant_LLL}-\ref{CLLL}) with the eigenvalue (\ref{epskern}). Another representation of such energy eigenstates can be found using the commutation relations of $Z$ and $Z^\dagger$:
\begin{align*}
\ket{\tilde{\psi}^{N}_M} &=  \sum^{M}_{m=0} \frac{(-1)^m}{\dsty m! N^m} 
 \prod_{k=1}^m (Z Z^\dagger-k N)  \ket{N-M,M,0...} .
\end{align*}
To compute the norm squared of (\ref{eq: unnormalized ladder states}), we keep in mind that $\ket{\tilde{\psi}^{N}_M}$ satisfies (\ref{Zkern}), which yields
$$
\braket{\tilde{\psi}^{N}_M|\tilde{\psi}^{N}_M} =b_0\braket{N-M,M,0...|\tilde{\psi}^{N}_M}.
$$
Using (\ref{eq: Zdagger action on O and 1}) and remembering that $b_0=1$ in (\ref{eq: unnormalized ladder states}), this can be written as
 $$
\braket{\tilde{\psi}^{N}_M|\tilde{\psi}^{N}_M} = \sum^M_{m=0} \frac{(-1)^m}{\dsty m! N^m}  \frac{M!}{(N-M)! } \frac{(N-(M-m))!}{(M-m)!}.
 $$
The right-hand side can be summed up in a closed form in terms of a generalized Laguerre polynomial:
  \begin{align}
\braket{\tilde{\psi}^{N}_M | \tilde{\psi}^{N}_M} = \frac{M!}{N^{M}} \, L_M^{-N-1}(-N).
\label{eq: norms as laguerre}
 \end{align}
We then define the normalized ladder state in the $Z^\dagger$-kernel of level $(N,M)$ with $2\leq M \leq N$:
  \begin{align}
\ket{\psi^{N}_M} = \frac{\ket{\tilde{\psi}^{N}_M}}{\sqrt{\braket{\tilde{\psi}^{N}_M|\tilde{\psi}^{N}_M}}}.
\label{laddernorm}
 \end{align}

It is instructive to compute the occupation numbers in the ladder states (\ref{laddernorm}). We shall assume large $M$ and $N$ and a fixed ratio $d^2 \equiv M/N < 1$. In this limit, as shown in Appendix \ref{App: Expectation values in ladder states},
\begin{align}
\bra{\psi^{N}_M} a_n^\dagger &a_n \ket{\psi^{N}_M}  \nonumber \\ \approx &\frac{N}{n!} \frac{e^{d^2-d}}{(1-d)} (n-(1-d^2))^2 (d-d^2)^n.
\label{eq: expectation value product text}
 \end{align}
This is exactly the same, under the identification $|p|=\sqrt{d-d^2}$ as the amplitude spectrum $|\al_n|^2$ of the classical stationary solutions of the LLL Hamiltonian given by (21) of \cite{BBCE} under the constraint that the classical expression for Z, which is (29) of \cite{BBCE}, vanishes. Of course, while the amplitudes of the classical stationary solution are reproduced by (\ref{laddernorm}), the classical evolution of the phases is not, since (\ref{laddernorm}) is an eigenstate of the quantum Hamiltonian. One should think of (\ref{laddernorm}) as an analog of energy eigenstates of a simple harmonic oscillator, whereas what one needs to fully relate to the classical evolution are analogs of coherent states. We shall construct such states explicitly in the next section, developing a thorough connection to the classical considerations of \cite{BBCE}.

 
\section{Coherent-like states and\\ connection to classical theory}\label{sec:coherent}

The ladder states given by (\ref{eq: unnormalized ladder states}) and (\ref{laddernorm}) discovered in the previous section have the property that any superposition of such states evolves periodically with a period proportional to $1/g$. This is very reminiscent of the periodicity properties of an explicit family of classical solutions of the LLL Hamiltonian constructed in \cite{BBCE}.
It is natural to look for specific superpositions of the ladder state that connect to the classical dynamics of
\cite{BBCE} as closely as possible. To this end, we define the following coherent-like states:
\begin{align}
\ket{\alpha,\beta} =  &e^{- | \alpha |^2/2}\sum_{N=0}^\infty (1+|\beta|^2)^{-N/2} \frac{\alpha^N}{\sqrt{N!}}  \label{eq: definition coherent state}\\ 
&\times \sum_{M=2}^N \beta^M \sqrt{ \frac{N!}{M! (N-M)!}} \ket{\psi^{N}_M}.\nonumber
\end{align}
We shall be mostly interested in the regime of large $|\alpha|$, and $|\beta|$ of order one. 
In this regime, the states defined above become semiclassical in the sense that the standard deviation of $a_n$ becomes small compared to the expectation value. This will be made manifest in our subsequent computations. (In all of our subsequent computations,
the mode number $n$ is kept fixed as $\left| \alpha \right|$ is taken large.)

The first thing we would like to analyze in relation to (\ref{eq: definition coherent state}) is the expectation
value $\bra{\alpha,\beta} a_n \ket{\alpha,\beta}$. The details of the computation are given in Appendix \ref{App: Expectation values in coherent states}. In a nutshell, the only nonvanishing matrix elements of the form  $\bra{\psi^{N'}_{M'}} a_n\ket{\psi^{N}_M}$ 
are for $N'=N-1$ and $M'=M-n$.
The sum over $M$ is then evaluated using the fact that the binomial distribution is sharply peaked around its maximum at large $N$, while the sum over $N$ reduces to 
\begin{equation}
\sum_{N=1}^\infty \frac{N\xi^N }{N!}=\xi e^\xi.
\label{a_sum}
\end{equation}
The end result is
\begin{align}
\bra{\alpha,\beta} a_n \ket{\alpha,\beta} &\approx  - \alpha   e^{-\frac{|p|^2}{2}} \frac{\left(n
    -\frac{1}{\left| \beta \right| ^2+1}\right)}{\sqrt{1-\sqrt{\frac{\left| \beta \right| ^2}{\left| \beta \right|
   ^2+1}}}}
    \frac{ p^{n}}{\sqrt{n!}} \nonumber \\
    &\equiv \left( \frac{a n}{p} +b \right) \frac{p^n}{\sqrt{n!}},
    \label{expectan}
\end{align}
where $a$ and $b$ can be read off by comparing the last line to the previous line, and
\begin{equation}
p = - \frac{\left|\beta \right|}{\beta^*} \sqrt{\sqrt{\frac{\left| \beta \right|^2}{1+\left| \beta \right|^2}}-\frac{\left| \beta \right|^2}{1+\left| \beta \right|^2}}
   \end{equation}
This expectation value thus agrees with the classical ansatz (21) in \cite{BBCE}, supporting our view that the ladder states are a quantum counterpart of the time-periodic classical solutions of \cite{BBCE}.

One can similarly evaluate the expectation values of the quantum operators $N,M$ and $Z$. 
Indeed, we give an explicit analysis in Appendix~\ref{App: Expectation values in coherent states} of the expectation values of $a_n^\dagger a_n$ in (\ref{eq: definition coherent state}). At large $ |\alpha |$, the result is
 \begin{align}
\bra{\alpha,\beta} a^\dagger_n a_n \ket{\alpha,\beta} &\approx \frac{| \alpha |^{2}}{n!}  \frac{e^{-|p|^2}
   \left(n-\frac{1}{\left| \beta \right| ^2+1}\right)^2 |p|^{2n}}{1-\sqrt{\frac{\left| \beta \right| ^2}{\left| \beta \right|
   ^2+1}}}
 \end{align}
where $\left| p \right| ^2=\sqrt{\frac{\left| \beta \right| ^2}{\left| \beta \right|
   ^2+1}}-\frac{\left| \beta \right| ^2}{\left| \beta \right| ^2+1}$. One then has the following expressions for the expectation values of $N$ and $M$:
\begin{align}
\sum_{n=0}^\infty \bra{\alpha,\beta} a_n^\dagger a_n \ket{\alpha,\beta} &= \left| \alpha \right|^2 \\
\sum_{n=0}^\infty \bra{\alpha,\beta}n  a_n^\dagger  a_n \ket{\alpha,\beta} &= \left| \alpha \right|^2 \frac{\left| \beta \right|^2}{1+\left| \beta \right|^2}.
\label{eq: normalization check}
\end{align}
Furthermore, $\bra{\alpha,\beta} Z \ket{\alpha,\beta} = 0$ since (\ref{eq: definition coherent state}) satisfy (\ref{Zkern}).
Note that, if one extracts $a$ and $b$ from (\ref{expectan}) and substitutes them to the classical expressions for the conserved quantities, given by (27-29) of \cite{BBCE}, one obtains the same values: $N=\left| \alpha \right|^2$, $M= \left| \alpha \right|^2\frac{\left| \beta \right|^2}{1+\left| \beta \right|^2}$, $Z=0$. We shall explain how to obtain more general states with nonzero values of $Z$ at the end of this section.

We now explore the time dependence of the states (\ref{eq: definition coherent state}) under the evolution defined by the LLL Hamiltonian (\ref{resonant_LLL}-\ref{CLLL}). In the language of the original system (\ref{Horg}), this corresponds to viewing the system in a reference frame rotating with the harmonic trap frequency, and in terms of the {\it slow time} $\tau=gt$. (In the formulas below giving time dependences, we shall simply use $t$ to refer to this slow time.)

We start with computing
\begin{equation}
\braket{ a_n }_t \equiv \bra{\alpha,\beta} e^{i\mathcal{H}_{LLL}t} a_n  e^{-i\mathcal{H}_{LLL}t} \ket{\alpha, \beta}.
\end{equation}
The computation is essentially identical to (\ref{expectan}), which is described in Appendix~\ref{App: Expectation values in coherent states}. The only nonvanishing matrix elements are still  
$\bra{\psi^{N-1}_{M-n}} a_n\ket{\psi^{N}_M}$, except that now they are multiplied with the phase factors $e^{i\om_{MN}t}$, where $\om_{MN}$ is expressed through the energies (\ref{epskern}):
\begin{align}
\omega_{MN}=&\,\eps^{(N-1,M-n)}-\eps^{(N,M)}\\
&\hspace{5mm}=\frac{(n-4)(N-1)+M}4.\nonumber
\end{align}
One then has to repeat the computations of Appendix~\ref{App: Expectation values in coherent states} taking account these phase factors, which is straightforward as the extra phases are linear in $N$, $M$ and $t$. The sum over $M$ is evaluated  as in Appendix~\ref{App: Expectation values in coherent states}.
Thereafter, the sum over $N$ is evaluated using  (\ref{a_sum})
with $\xi = \left| \alpha \right|^2 e^{ \frac{i t}{4}\left(\frac{\left| \beta \right| ^2}{ 1+\left| \beta \right| ^2}+n-4\right)}$, which gives
\begin{align}
\braket{a_n}_t = \braket{a_n}_0  e^{\frac{i t}{4} \frac{\left| \beta \right| ^2}{ 1+\left| \beta \right| ^2}} e^{\left| \alpha \right|^2 \left(e^{ \frac{i t}{4}\left(\frac{\left| \beta \right| ^2}{ 1+\left| \beta \right| ^2}+n-4\right)}-1\right)},
\label{eq: time evolution a_n}
\end{align}
where $\braket{a_n}_0$ is the same as in (\ref{expectan}).

In order to study the classical limit, one needs to restore the factors of $\hbar$. In the classical limit $\hbar\to 0$, the canonical coordinates $x_n\sim\sqrt{\hbar}(a_n+a^\dagger_n)$ should stay finite, together with their conjugate canonical momenta, and the LLL Hamiltonian (\ref{resonant_LLL}) must stay finite in terms of these canonical coordinates and momenta. This means that the LLL Hamiltonian (\ref{resonant_LLL}) must be prefaced with $\hbar^2$. Together with the standard $\hbar$ in front of the time derivative in the Schr\"odinger equation, this means that reinserting $\hbar$ effectively amounts to replacing $t$ by $\hbar t$ in our expressions. The classical limit is then obtained by taking $\hbar \rightarrow 0$ while keeping $\left| \braket{a_n} \right|^2 \hbar$ fixed. The quantum dynamics is expected to reproduce classical features as long as the variance of the expectation value of $a_n$ in the coherent state is small with respect to the modulus squared of its expectation value. By computations closely retracing the steps of Appendix~\ref{App: Expectation values in coherent states}, we find
\begin{align}
&\braket{ a^\dagger_n a_n }_t - \left|\braket{  a_n }_t \right|^2\nonumber \\ 
&=  | \braket{a_n}_0 | ^2 \left(1-e^{2| \alpha |^{2}\left( \cos\left(\frac{1}{4}\frac{\left| \beta \right| ^2}{ 1+\left| \beta \right| ^2}+\frac{n}{4}-1\right)\hbar t-1 \right)}\right) \nonumber\\
   &\approx | \braket{a_n}_0 | ^2 | \alpha |^{2} \left(\frac{1}{4}\frac{\left| \beta \right| ^2}{ 1+\left| \beta \right| ^2}+\frac{n}{4}-1\right)^2 \hbar^2 t^2.
\end{align}
Therefore,
\begin{align}
\frac{\braket{ a^\dagger_n a_n }_t - \left|\braket{  a_n }_t \right|^2}{\left|\braket{  a_n }_t \right|^2}   \sim | \alpha |^{2} \hbar^2 t^2.
\end{align}
At $t\sim 1$, $|\alpha|^2\sim 1/\hbar$ and $\hbar\to 0$, this expression vanishes, defining a classical state.
One can similarly reinsert $\hbar$ into \eqref{eq: time evolution a_n} and take the $\hbar\to 0$ limit to recover the classical LLL solutions of \cite{BBCE}. 

A remark is in order on our usage of perturbation theory. In the previous section, the number of particles $N$ was kept fixed and $g$ taken to be small. In this section, however, for comparing with the regime of Bose-Einstein condensation, we take the limit of large $N$. One may wonder whether having a large parameter $N$ alongside the small parameter $g$ upsets the operation of the perturbation theory. While it would be difficult to accurately analyze the higher orders of quantum-mechanical perturbation theory, our main goal in considering large $N$ is making contact with the classical (mean-field) theory, and in that setting, the structure of the perturbative expansion is well-understood. Indeed, the Gross-Pitaevskii equation possesses simple scaling properties, which make it possible to absorb the coupling parameter $g$ into the normalization of the wavefunction (or the other way around). There is then only one effective coupling parameter, $gN$, and `small coupling' in the context of Bose-Einstein condensates means precisely that this parameter is small. In this regime, the higher-order corrections to perturbative treatments are expected to be small as long as $gN$ is small. Indeed, proofs are available in the mathematical literature \cite{continuous} that the resonant approximation, which in the LLL sector is just given by the classical version of the Hamiltonian (\ref{resonant_LLL}-\ref{CLLL}), describes the dynamics of the Gross-Pitaevskii equation arbitrarily well over long times when the coupling parameter is small. It is natural to expect that the expansion parameter in quantum theory is similarly $gN$. (This is indeed the case for the lowest order of perturbation theory we have considered, as the unperturbed energy $M$, of the same order as $N$, gets corrected by a contribution of order $gN^2$, since the shifts are distributed between 0 and the maximal shift corresponding to (\ref{N000}), which is of this order.)

Classical states of the form (21) in \cite{BBCE} define a three-dimensional invariant manifold. 
The coherent states \eqref{eq: definition coherent state} are constructed with two free parameters $\alpha$ and $\beta$ which tune $N$ and $M$, but are constrained to satisfy $Z^\dagger\ket{\alpha,\beta}=0$ as they are explicitly built out of $Z^\dagger$-kernel eigenstates.
The third free parameter can be reinstated by applying the unitary operator $e^{q^*Z-qZ^\dagger}$ to the coherent states (\ref{eq: definition coherent state}):
\begin{equation}
\ket{\alpha,\beta,q} \equiv e^{q^*Z-qZ^\dagger}  \ket{\alpha,\beta}.
\label{albeq}
\end{equation} 
This is in parallel to the classical story of \cite{BBCE}, where the $Z$-transformation can be applied to solutions with $Z=0$, to obtain other time-periodic solutions. 
After some algebra, one can show (see Appendix \ref{App: Construction of coherent-like states with nonzero momentum}) that these states reproduce the classical transformations given by (35) of \cite{BBCE} for $a$, $b$ and $p$,
\begin{align}
\bra{\alpha,\beta,q} &a_n \ket{\alpha,\beta,q}  \\ 
&= \frac{e^{-pq-\frac{1}{2}\left|q \right|^2}}{\sqrt{n!}} \left(\frac{a n}{p+q^*}+b- a q \right) (p+q^*)^n .\nonumber
\end{align}


\section{Discussion}

Motivated by time-periodic behaviors with periods of order $1/g$ in the classical (mean-field) theory of Bose-Einstein condensates at the lowest Landau level \cite{BBCE}, we have analyzed the fine splitting of the energy levels of the corresponding quantum problem at order $g$. In the fine structure emanating from an integer unperturbed LLL level with $N$ particles, $M$ units of energy and $M$ units of angular momentum, we have discovered ladders of equispaced energy eigenstates (with energy distances proportional to $g$). The energy shifts of these states (relatively to the unperturbed level of noninteracting bosons from which they split off), in units of $g$, are given by (\ref{epsladder}), while their explicit (unnormalized) wavefunctions are 
\beq
\ket{\psi^{(N,M)}_m}=Z^{m}\ket{\tilde{\psi}^N_{M-m}},
\eeq
with $\ket{\tilde{\psi}^N_m}$ given by (\ref{eq: unnormalized ladder states}) and $Z$ given by (\ref{ZLLL}). Coherent-like combinations of these states can be constructed in the form (\ref{eq: definition coherent state}) and (\ref{albeq}). In an appropriate semiclassical regime, these states closely approximate the classical time-periodic dynamics discussed in \cite{BBCE}. We should emphasize, however, that the amount of time-periodicity one discovers in the quantum theory is much greater than what the classical theory would naively suggest. Indeed, the entire family of special time-periodic classical solutions in \cite{BBCE} is parametrized by three complex numbers. One the other hand, as manifest from our discussion, the corresponding time periodicities in the quantum theory are present for arbitrary superpositions of the giant family of states we construct.

Time-periodic dynamics of various sorts has been discussed for Bose-Einstein condensates in harmonic traps in the past literature \cite{BBCE, quasiint1,quasiint2,tri}. Such abundant time-periodic behaviors are often seen as signals of `quasi-integrable' dynamics of the Gross-Pitaevskii equation in low dimensions, and have been discussed, in particular, from the standpoint of non-ergodicity \cite{quasiint2}. Our present considerations provide new insights on this issue by showing how such time-periodic behaviors extend and even become enriched outside the mean field regime described by the Gross-Pitaevskii equation.

While our main motivations were in exploring the structures of the spectra in the quantum theory, our considerations shed light on the corresponding classical (mean field) dynamics, which is unusual, as one normally thinks of the quantum theory as being much more complicated than its classical counterpart. Indeed, since
the bound (\ref{Bbound}) is established by analyzing the structure of (\ref{eq: resonant definition B}), a similar statement should hold in the classical theory as an inequality
\beq
\mathcal{H}_{LLL}\ge\frac{N^2}{2}-\frac{NM-|Z|^2}4.
\label{classineq}
\eeq
It turns out that the time-periodic classical solutions of \cite{BBCE} precisely saturate this inequality, in an immediate relation to the eigenvectors (\ref{zerov}). They thus lie at the bottom of a valley in phase space, which gives a natural explanation for the consistency of the corresponding ansatz, a puzzle since it was established in \cite{BBCE}.
We remark that it would be very difficult to guess the inequality (\ref{classineq}) from purely classical reasoning, while its analog in the corresponding quantum theory is strongly suggested by the patterns in the energy spectra visible from straightforward numerics in the spirit of \cite{quantres,split}, which is how we have arrived at it in practice. Quantization, in this case, brings in surprising benefits.

Our study has focused on theoretical properties of bosons with contact interactions described by the Hamiltonian (\ref{Horg}). While not pursued here, it is  intriguing to look for real-life manifestations
of the quantum states we construct, in particular in cold atomic gases that can be described well by the Hamiltonian (\ref{Horg}). A key question in this regard is how to manufacture our states in a lab. While devising concrete experimental setups is beyond our current goals, it is a welcome feature that the ladder states are ground states of the modified Hamiltonian (\ref{eq: definition B}). A familiar underlying principle for creating states in a lab is to find a Hamiltonian for which the desired states are at the lowest energy level, modify the trapping conditions to reproduce this Hamiltonian, and let the system settle to the ground state. We note the recent work on experimental realization of prescribed initial configurations via similar principles (albeit in the regime of Bose-Einstein condensation) in \cite{tri}, also strongly motivated by questions of time-periodicity, breathing modes and breathers.

There is a number of ways our considerations may generalize. The LLL Hamiltonian (\ref{resonant_LLL}-\ref{CLLL}) is a particularly simple representative of a large class of resonant Hamiltonians constructed in \cite{AO}, admitting special time-periodic classical solutions. Such Hamiltonians naturally arise in weakly nonlinear classical theory for Bose-Einstein condensates \cite{BBCE2}, for Hartree-type equations \cite{BEF} and for relativistic analogs of these problems \cite{CF, BEL}, which are of interest in gravitational and high-energy physics. The generalities of how such solvable weakly nonlinear dynamics may arise as a controlled approximation to PDEs have been spelled out in \cite{breathing}. This, in particular, implies that such analytic structures are to be expected for the Gross-Pitaevskii equation with a harmonic potential in one dimension, and, within the maximally rotating sector, in three dimensions, generalizing the two-dimensional considerations of our paper. It is legitimate to expect that the corresponding quantum systems will display features similar to what we have described here, and can be analyzed by similar means.\footnote{We also point out potential connections to the topic of quantum anomaly in the Pitaevskii-Rosch breathing mode, treated in \cite{qua}.
This feature emerges in the context of the same system (\ref{Horg}) that we have studied here, but would require going outside the LLL sector, since the Pitaevskii-Rosch breathing mode plays no role in the dynamics of the LLL sector and has vanishing matrix elements between any two LLL states.} The solvable time-periodic features often arise from truncating the corresponding theories to subsets of modes (spherically symmetric or maximally rotating, for example) in the weakly nonlinear regime. Some of such truncations (say, the maximally rotating sectors) will be inherited by the corresponding quantum theory, as happened to the LLL states in our present treatment. In other cases, the truncation may no longer be strictly speaking valid in the quantum theory, but it should still be approximately valid for states of high energy, where classical dynamics is approached. In such situations, one expects that ladders in the fine structure splitting, analogous to what we have seen for the LLL Hamiltonian, will emerge asymptotically in the fine structure of sufficiently high unperturbed energy levels. We intend to investigate some of these questions elsewhere \cite{prep}.


\begin{acknowledgments}

We thank Ben Craps and Surbhi Khetrapal for collaboration on related subjects.
This research has been supported by FWO-Vlaanderen (projects G044016N and G006918N), by Vrije Universiteit Brussel through the Strategic Research Program ``High-Energy Physics,'' and by CUniverse research promotion project (CUAASC) at Chulalongkorn University. M.~D.~C. is supported by a PhD fellowship from the Research Foundation Flanders (FWO). O.~E. has benefited from his stay at the Jagiellonian University in Krakow during which part of this work was developed. Support of the Polish National Science Centre through grant number 2017/26/A/ST2/00530 and personal hospitality of 
Piotr and Magda Bizo\'n are gratefully acknowledged.
\end{acknowledgments}

\onecolumngrid

\appendix

\section{Expectation values in the ladder states}
\label{App: Expectation values in ladder states}
We shall compute the expectation values $a_n^\dagger a_n$ and $a_n$ in ladder states \eqref{laddernorm} and show that in the large $N$, $M$ limit with fixed ratio $d^2 \equiv M/N$
$$\braket{\psi^N_M|a_n^\dagger a_n|\psi^N_M} \approx |\bra{\psi^{N-1}_{M-n}} a_n \ket{\psi^{N}_M}|^2.$$
We start by listing a few useful identities. 
In general, commuting $a_n$ past $Z^m$, which are contained in $\ket{\psi^N_M}$ as per (\ref{eq: unnormalized ladder states}), gives
\begin{equation}
a_n Z^m = \sum_{k=max(0,n-m)}^n Z^{k+m-n}\, a_k \, \sqrt{\frac{n!}{k!}}\, \frac{m!}{(n-k)!(m-n+k)!}.
\label{eq: identity a through Z}
\end{equation}
Similarly one can commute $Z^{\dagger l}$ past an annihilation operator, which reduces to the following simple form on states $\left\lbrace \ket{\psi} \right\rbrace$ in the kernel of $Z^\dagger$, 
\begin{equation}
Z^{\dagger l}a_m \ket{\psi} = (-1)^l \sqrt{\frac{(m+l)!}{m!}} a_{m+l} \ket{\psi}.
\label{eq: Zdam}
\end{equation}
Finally, we will make use of the following identify \cite{monomials}, 
\begin{align}
Z^{\dagger m} Z^{m'} = \sum_{k=0}^{min(m,m')} \frac{N^k m! m'!}{k! (m-k)! (m'-k)!} Z^{m'-k}Z^{\dagger m-k}.
\label{eq: commutation products of Z's}
\end{align}

First, we will argue that the state $a_n \ket{\psi^N_M}$ only contains ladder states when decomposed in eigenstates of $\mathcal{H}_{LLL}$ at level $(N-1,M-n)$. To this end, we write $\ket{\psi^N_M}$ using \eqref{eq: unnormalized ladder states} and drag the operator $a_n$ through the products of $Z$-operators using \eqref{eq: identity a through Z}. In this case, any $a_k$ with $k\ge 2$ annihilates $Z^{\dagger m}|N-M,M,0...\rangle$ contained in $\ket{\psi^N_M}$, as seen from \eqref{eq: Zdagger action on O and 1}, effectively reducing (\ref{eq: identity a through Z}) to two terms at most. The action of $a_0$ or $a_1$ on  $Z^{\dagger m}|N-M,M,0...\rangle$ is straightforward, obtained by acting with these operators on (\ref{eq: Zdagger action on O and 1}). As a consequence, the products of operators $Z$ that were pulled out now act on states proportional to $|i,j,0...\rangle$, for some $i$ and $j$, which lie in the kernel of $B$. Because $B$ and $Z$ commute, the state $a_n \ket{\psi^N_M}$ is also annihilated by $B$ and can therefore only contain ladder states in the $(N-1,M-n)$-block,
\begin{equation}
a_n \ket{\psi^{N}_M} = \sum_{l=0}^{M-n} \frac{\bra{\psi^{N-1}_{M-n-l}} Z^{\dagger l} a_n\ket{\psi^{N}_M} }{\bra{\psi^{N-1}_{M-n-l}} Z^{\dagger l}Z^l \ket{\psi^{N-1}_{M-n-l}}}Z^l \ket{\psi^{N-1}_{M-n-l}}.
\end{equation}
The expectation value of $a^\dagger_n a_n$ in ladder states therefore reduces to
\begin{align}
\bra{\psi^{N}_{M}}a^\dagger_n a_n\ket{\psi^{N}_M} &= \sum^{M-n}_{l=0} \left|\bra{\psi^{N-1}_{M-n-l}}Z^{\dagger l} a_n\ket{\psi^{N}_M}\right|^2 \label{eq: formula amplitude spectrum} \\
&=\sum^{M-n}_{l=0} \frac{\left|\bra{\tilde{\psi}^{N-1}_{M-n-l}}Z^{\dagger l} a_n\ket{\tilde{\psi}^{N}_M}\right|^2}{\braket{\tilde{\psi}^{N}_{M}|\tilde{\psi}^{N}_{M}} \bra{\tilde{\psi}^{N-1}_{M-n-l}} Z^{\dagger l} Z^{ l}\ket{\tilde{\psi}^{N-1}_{M-n-l}}} \nonumber .
\end{align}
The denominator is easily determined using \eqref{eq: norms as laguerre} and \eqref{eq: commutation products of Z's}. We compute the numerator by applying \eqref{eq: Zdam} followed by \eqref{eq: identity a through Z} in the decomposition \eqref{eq: unnormalized ladder states} of $\ket{\tilde{\psi}^{N}_M}$. As before, only terms with $k < 2$ produce nonvanishing contributions, while any factor of $Z$ that was pulled out by this action annihilates $\bra{\tilde{\psi}^{N-1}_{M-n-l}}$. After that, one can expand the bra as \eqref{eq: unnormalized ladder states} and use the identity \eqref{eq: Zdagger action on O and 1} twice. Completing the algebra, one finds
\begin{multline}
\bra{\tilde{\psi}^{N-1}_{M-n-l}}Z^{\dagger l} a_n\ket{\tilde{\psi}^{N}_M} = \frac{(-1)^n}{\sqrt{n!}}\frac{\sqrt{M!(M-n-l)!(N-1-M+n+l)!}}{N^{n+l}(N-1)^{M-n-l}\sqrt{(N-M)!}} L^{-N}_{M-n-l}(1-N) \\
\times \left(N-M-(n+l)(N-1) \right),
\end{multline}
so that
\beq
\left|\bra{\psi^{N-1}_{M-n-l}}Z^{\dagger l} a_n\ket{\psi^{N}_M}\right|^2 = \frac{N^M(N-1-M+n+l)!}{N^{2n+2l}(N-1)^{M-n}(N-M)!} \frac{L^{-N}_{M-n-l}(1-N)}{L^{-N-1}_{M}(-N)}  \frac{\left(N-M-(n+l)(N-1) \right)^2}{n! l!} .
\eeq

Using the limiting behavior of Laguerre polynomials of the form $L_{n}^{\alpha_n}(z n)$ with $ \lim_{n \rightarrow \infty} -\alpha_n/n > 1$, derived in \cite{McLaughlin,Wong}, one can show that for large $M$ and $N$, with fixed $M/N=d^2$,
\begin{equation}
\frac{L_{M-n-l}^{-N}(1-N)}{L_M^{-N-1}(-N)} \approx (1+d) \, e^{-d}\left(\frac{1}{1+1/d}\right)^{n+l}.
\label{eq: fraction of laguerre polynomials}
\end{equation}
In particular, (7.5) of \cite{McLaughlin} describes the behavior of generalized Laguerre polynomials with a large parameter in the region of the complex plane of interest for us, while an explicit representation of the various functions appearing in this expression are given in (3.1) and (3.2) of \cite{McLaughlin} and (4.14) and (4.15) of \cite{Wong}. 

From this, we can conclude that the summand with $l=0$ dominates the sum in \eqref{eq: formula amplitude spectrum} at large $N$ and $M$ with fixed $d$, because
\begin{equation}
\frac{\left|\bra{\psi^{N-1}_{M-n-l}}Z^{\dagger l} a_n\ket{\psi^{N}_M}\right|^2}{\left|\bra{\psi^{N-1}_{M-n}} a_n\ket{\psi^{N}_M}\right|^2
} \sim \frac{(N-M)^l}{N^{2l}} \sim \frac{1}{N^l} \rightarrow 0,
\end{equation}
for all $l>0$. 
Therefore, 
\begin{equation}
\bra{\psi^{N}_{M}}a^\dagger_n a_n\ket{\psi^{N}_M} \approx \left|\bra{\psi^{N-1}_{M-n}} a_n\ket{\psi^{N}_M}\right|^2
\label{eq: expectation value product}
\end{equation}
with 
\begin{align}
\bra{\psi^{N-1}_{M-n}} a_n \ket{\psi^{N}_M} = & \frac{(-1)^n N^{M/2}}{N^n (N-1)^{(M-n)/2}}\sqrt{\frac{(N-1-M+n)!}{(N-M)!}}  \frac{(N-M-n(N-1))}{\sqrt{n!}}  \sqrt{\frac{L_{M-n}^{-N}(1-N)}{L_M^{-N-1}(-N)}} \nonumber\\
\approx & \frac{(-1)^n }{N^{n/2}}\sqrt{(N-M)^{n-1}} \, \frac{(N-M-n(N-1))}{\sqrt{n!}} \sqrt{\left(1+\sqrt{\frac{M}{N}}\right)e^{\frac{M}{N}-\sqrt{\frac{M}{N}}}\left(\frac{1}{1+\sqrt{\frac{N}{M}}}\right)^n} \nonumber\\
\approx & \frac{(-1)^n \sqrt{N}}{\left(1-d\right)^{1/2}} \, \frac{(1-d^2-n)}{\sqrt{n!}} e^{\frac{d^2-d}{2}}\left(d-d^2\right)^{n/2},
\label{eq: exp value single terms}
\end{align}
in the limit of large $N$ and $M$, keeping the ratio $d^2 = M/N$ fixed.
As a cross-check, those results in fact agree well with numerical computations of such expectation values.
It also has the right normalization of the amplitude spectrum:
\begin{align*}
\bra{\psi^{N}_M}\sum_{n=0}^M a_n^\dagger a_n\ket{\psi^{N}_M} &= N \\
\bra{\psi^{N}_M}\sum_{n=0}^M n a_n^\dagger a_n\ket{\psi^{N}_M} &= M,
\end{align*}
for $M,N \gg 1$.

\section{Expectation values in the coherent states}
\label{App: Expectation values in coherent states}
 Given the expectation values in ladder states \eqref{eq: expectation value product} and \eqref{eq: exp value single terms}, one can evaluate similar expectation values in the coherent-like states \eqref{eq: definition coherent state}.
We first compute the expectation value of $a_n$ in $\ket{\alpha,\beta}$:
\begin{align}
\bra{\alpha,\beta} a_n \ket{\alpha,\beta} = &e^{- | \alpha |^2} \sum_{N=1}^\infty \frac{| \alpha |^{2 N} \sqrt{N}/ \alpha^{*}}{N!} (1+|\beta|^2)^{-N+1/2} \nonumber \\ \times &\sum_{M=n}^N \frac{|\beta|^{2 M}}{\beta^{*n}} \sqrt{\frac{N!}{M! (N-M)!}}\sqrt{\frac{(N-1)!}{(M-n)! (N-1-(M-n))!}} 
\bra{\psi^{N-1}_{M-n}} a_n \ket{\psi^{N}_M}.
\end{align}
(Note that the second sum has $N-1$ as its upper bound in the case $n=0$.)
For $ \left| \alpha \right| \gg 1$ the first sum will be dominated by terms with $N \sim | \alpha |^2$. Because of the presence of the binomial coefficient in the second sum, for large $N$, the leading terms will be centered around $M \sim N|\beta|^2/(1+|\beta|^2)$, which means that we are in the regime where \eqref{eq: exp value single terms} is a good approximation. Keeping in mind that
\begin{align*}
\sqrt{\frac{N!}{M! (N-M)!}} \sqrt{\frac{(N-1)!}{(M-n)! (N-1-(M-n))!}} \approx \frac{N!}{M! (N-M)!} \sqrt{\frac{(N-M)}{N}}\left(\frac{M}{N-M}\right)^{n/2},
\end{align*}
the sum over $M$ becomes schematically of the form $\sum_M x^M {N \choose M}\,\, f(M/N)$, where $f$ is a smooth function.
The distribution $x^M{N \choose M}$ has mean ${Nx}/(1+x)$ and a standard deviation of order $\sqrt{N}$. Within the standard deviation, $f$ essentially does not vary, provided that $N$ is large, and can be taken outside the sum as $f(x/(1+x))$. 
Although this approximation is only valid in the large $N$ limit, if one takes $\left| \alpha \right|$ to be large, only terms with $N$ large will contribute in the $N$-sum, so the approximation can be legitimately employed. 
In conclusion, one obtains
\begin{align*}
\bra{\alpha,\beta} a_n \ket{\alpha,\beta} &= e^{- | \alpha |^2} \sum_{N=1}^\infty \frac{| \alpha |^{2 N} N}{N!\alpha^{*}}   \frac{e^{\frac{1}{2}\frac{\left| \beta \right|
   ^2}{\left| \beta \right| ^2+1}-\frac{1}{2}  \sqrt{\frac{\left| \beta \right|^2}{\left| \beta \right| ^2+1}}} }{\sqrt{1-\sqrt{\frac{\left| \beta \right| ^2}{\left| \beta \right|
   ^2+1}}}}  \frac{(-1)^{n}}{\beta^{*n}}
    \left(\frac{\left| \beta \right|^2/(\left| \beta \right| ^2+1)}{\sqrt{\frac{1 + \left| \beta \right| ^2}{\left|
   \beta \right| ^2}}+1 }\right)^{n/2} \frac{\left(\frac{1}{\left| \beta \right| ^2+1}-n\right)}{\sqrt{n!}} \\
   & \approx  - \alpha   e^{-\frac{|p|^2}{2}} \frac{1}{\sqrt{1-\sqrt{\frac{\left| \beta \right| ^2}{\left| \beta \right|
   ^2+1}}}}
    \frac{ p^{n}}{\sqrt{n!}} \left(n
    -\frac{1}{\left| \beta \right| ^2+1}\right)\\
    &\equiv \left( \frac{a n}{p} +b \right) \frac{p^n}{\sqrt{n!}} 
\end{align*}
with
\begin{equation}
p = - \frac{\left|\beta \right|}{\beta^*} \sqrt{\sqrt{\frac{\left| \beta \right|^2}{1+\left| \beta \right|^2}}-\frac{\left| \beta \right|^2}{1+\left| \beta \right|^2}}.
 \end{equation}
 \newline
 \newline
 Repeating these steps for the expectation value of $a^\dagger_n a_n$ in $\ket{\alpha,\beta}$
   \begin{align*}
\bra{\alpha,\beta} a^\dagger_n a_n &\ket{\alpha,\beta} = e^{- | \alpha |^2} \sum_{N=0}^\infty \frac{| \alpha |^{2 N}}{N!} (1+|\beta|^2)^{-N}  \sum_{M=n}^N |\beta|^{2 M}  \frac{N!}{M! (N-M)!}
\bra{\psi^{N}_{M}} a^\dagger_na_n \ket{\psi^{N}_M},
\end{align*}
and taking into account \eqref{eq: expectation value product}, one finds:
   \begin{align}
\bra{\alpha,\beta} a^\dagger_n a_n \ket{\alpha,\beta} &\approx \frac{e^{- | \alpha |^2 }}{n!}  \sum_{N=1}^\infty \frac{| \alpha |^{2 N} N}{N!} \frac{e^{-|p|^2}
   \left(n-\frac{1}{\left| \beta \right| ^2+1}\right)^2 
   |p|^{2n}}{1-\sqrt{\frac{\left| \beta \right| ^2}{\beta
   ^2+1}}} \nonumber\\
   &\approx \frac{| \alpha |^{2}}{n!}  \frac{e^{-|p|^2}
   \left(n-\frac{1}{\left| \beta \right| ^2+1}\right)^2 
   |p|^{2n}}{1-\sqrt{\frac{\left| \beta \right| ^2}{\left| \beta \right|
   ^2+1}}},
 \end{align}
 with $\left| p \right| ^2= \sqrt{\frac{\left| \beta \right| ^2}{\left| \beta \right|
   ^2+1}}-\frac{\left| \beta \right| ^2}{\left| \beta \right| ^2+1}$. 
   
\section{Construction of coherent-like states with nonzero $Z$}   
\label{App: Construction of coherent-like states with nonzero momentum}   
We examine the expectation value of $a_n$ in the states (\ref{albeq}).
From the Baker-Campbell-Hausdorff identity,
\begin{equation}
e^{q^*Z-qZ^\dagger}  = e^{q^*Z}e^{-qZ^\dagger}e^{-\frac{1}{2} \left| q \right|^2 N},
\end{equation}
and \eqref{eq: identity a through Z}, we get
\begin{equation}
a_n e^{q^*Z} = e^{q^*Z} \sum_{m=0}^n \frac{(q^*)^{n-m}}{(n-m)!} \sqrt{\frac{n!}{m!}} a_m.
\end{equation}
Moreover, \eqref{eq: Zdam} implies 
\begin{equation}
e^{ qZ^\dagger}a_m \ket{\psi} = \sum_{l=0}^\infty \frac{(- q)^l}{l!}\sqrt{\frac{(m+l)!}{m!}} a_{m+l} \ket{\psi}.
\end{equation}
Therefore,
\begin{align}
\bra{\alpha,\beta,q} a_n \ket{\alpha,\beta,q} &= 
\bra{\alpha,\beta} e^{-q^*Z}e^{qZ^\dagger}e^{-\frac{1}{2}\left| q \right|^2 N} a_n e^{q^*Z}e^{-qZ^\dagger}e^{-\frac{1}{2}\left| q \right|^2 N}\ket{\alpha,\beta}  \nonumber \\
&= \bra{\alpha,\beta}e^{qZ^\dagger}e^{-\frac{1}{2}\left| q \right|^2 N} a_n e^{q^*Z}e^{-\frac{1}{2}\left| q \right|^2 N}\ket{\alpha,\beta}  \nonumber \\
&= \bra{\alpha,\beta}e^{qZ^\dagger}e^{-\left| q \right|^2 N-\frac{1}{2}\left| q \right|^2} a_n e^{q^*Z}\ket{\alpha,\beta}  \nonumber \\
&= \sum_{m=0}^n \frac{(q^*)^{n-m}}{(n-m)!} \sqrt{\frac{n!}{m!}}\bra{\alpha,\beta}e^{qZ^\dagger}e^{-\left| q \right|^2 N-\frac{1}{2}\left| q \right|^2} e^{q^*Z} a_m \ket{\alpha,\beta}  \nonumber \\
&= \sum_{m=0}^n \frac{(q^*)^{n-m}}{(n-m)!} \sqrt{\frac{n!}{m!}} e^{-\frac{1}{2}\left| q \right|^2} \bra{\alpha,\beta} e^{q^*Z} e^{qZ^\dagger} a_m \ket{\alpha,\beta}  \nonumber \\
&= \sum_{m=0}^n \frac{(q^*)^{n-m}}{(n-m)!} \sqrt{\frac{n!}{m!}} e^{-\frac{1}{2}\left| q \right|^2} \bra{\alpha,\beta} e^{qZ^\dagger} a_m \ket{\alpha,\beta}  \nonumber \\
&= \frac{e^{-\frac{1}{2}\left| q \right|^2}}{\sqrt{n!}}\sum_{l=0}^{\infty} \frac{(-q)^l}{l!} \sum_{m=0}^n (q^*)^{n-m} \frac{n!}{m!(n-m)!} \sqrt{(m+l)!}  \bra{\alpha,\beta}a_{m+l} \ket{\alpha,\beta}  \nonumber \\
&= \frac{e^{-\frac{1}{2}\left| q \right|^2}}{\sqrt{n!}}\sum_{l=0}^{\infty} \frac{(-q)^l}{l!} \sum_{m=0}^n (q^*)^{n-m} \frac{n!}{m!(n-m)!} \left(\frac{a(m+l)}{p}+b\right) p^{m+l}  \nonumber \\
&= \frac{e^{-\frac{1}{2}\left| q \right|^2-pq}}{\sqrt{n!}}\left(\frac{a n}{p+q^*}+b-a q\right)\left(p+q^*\right)^n 
\end{align}
where we used
\begin{equation}
a_n e^{-\frac{\left| q \right|^2}{2} N} = e^{-\frac{\left| q \right|^2}{2} (N+1)} a_n
\end{equation}
in going from the second to the third line, and the Baker-Campbell-Hausdorff identity in going from the fourth to the fifth line.

\twocolumngrid

\end{document}